\documentclass[aps,prd,onecolumn,a4paper,showpacs,nofootinbib,eqsecnum]{revtex4}

\usepackage{amssymb,amsmath,latexsym,mathrsfs}
\usepackage{graphicx}
\usepackage{epsfig}
\usepackage{varioref,xr-hyper}
\usepackage[hypertex,draft=false,verbose=false,debug=false
            hyperindex=true,hypertexnames=true]{hyperref}

\newcommand{\hide}[1]{{}}
\newcommand{\mr}[1]{\mathrm{#1}}
\newcommand{\si}[1]{{\scriptscriptstyle{#1}}}
\def\Hilb{\mathscr{H}}  
\def\Leb{\mathscr{L}}   

\newcommand{\cst}{\mathrm{cst}}

\newcommand{\adj}[1]{{#1}^\dagger}
\newcommand{\com}[2]{\left[ #1, #2 \right]}

\newcommand{\ket}[1]{\left| #1 \right\rangle}
\newcommand{\scal}[2]{(#1,#2)}

\newcommand{\ads}{\ensuremath{\mathrm{AdS}_{\si{5}}\,}}

\newcommand{\kappafive}{\kappa_{{_5}}}
\newcommand{\kappareduct}{\kappa_{{_4}}}
\newcommand{\tension}{\mathcal{T}}

\newcommand{\bb}{{\si{\bullet}}}
\newcommand{\ve}[1]{\boldsymbol{#1}}
\newcommand{\p}{\partial}
\newcommand{\ud}{\mathrm{d}}
\newcommand{\ub}{\mathrm{b}}

\newcommand{\yreg}{y_{{\mathrm{r}}}}
\newcommand{\ybr}{y_\ub}
\newcommand{\dybr}{\dot{y}_\ub}

\newcommand{\ybrR}{\widehat{y}_\ub}        
\newcommand{\tbr}{t_\ub}
\newcommand{\dtbr}{\dot{t}_\ub}

\newcommand{\Xbr}{X_\ub}

\newcommand{\Hb}{h_{\bb}}
\newcommand{\QHb}{\hat{h}_{\bb}}
\newcommand{\dHb}{\dot{h}_{\bb}}
\newcommand{\ddHb}{\ddot{h}_{\bb}}
\newcommand{\Pib}{\pi_{\bb}}
\newcommand{\QPib}{\hat{\pi}_{\bb}}

\newcommand{\PiTb}{\Pi^{\si{(T)}}_{\bb}}
\newcommand{\TKK} {\mathrm{m}}            
\newcommand{\TKKR}{\widehat{\mathrm{m}}}  
\newcommand{\TNa} {A}                     
\newcommand{\TNb} {B}                     
\newcommand{\TF}  {\mathrm{q}}            
\newcommand{\TTF} {\tilde{\mathrm{q}}}    
\newcommand{\TQF} {\hat{\mathrm{q}}}      
\newcommand{\TM}  {\mathrm{p}}            
\newcommand{\TQM} {\hat{\mathrm{p}}}      
\newcommand{\TQc} {\hat{a}^\dagger}       
\newcommand{\TQa} {\hat{a}}               

\newcommand{\HH}{\mathcal{H}}
\newcommand{\LL}{\mathcal{L}}

\newcommand{\vv}{\mathrm{v}}

\newcommand{\ga}{\gamma}
\newcommand{\ep}{\epsilon}
\newcommand{\dd}{\partial}
\newcommand{\ra}{\rightarrow}

\renewcommand{\ln}[1]{\,\mathrm{ln}\left( #1 \right)}
\renewcommand{\exp}[1]{\,\mathrm{exp}\left( #1 \right)}

\def\lsim{\;\raise 0.4ex\hbox{$<$}\kern -0.8em\lower 0.62 ex\hbox{$\sim$}\;}
\def\gsim{\;\raise 0.4ex\hbox{$>$}\kern -0.7em\lower 0.62 ex\hbox{$\sim$}\;}

\begin{document}

\title{On graviton production in braneworld cosmology}
\author{Cyril Cartier}
\email{cyril.cartier@physics.unige.ch}
\author{Ruth Durrer}
\email{ruth.durrer@physics.unige.ch}
\author{Marcus Ruser}
\email{marcus.ruser@physics.unige.ch}
\affiliation{D\'epartement de Physique Th\'eorique, Universit\'e de
Gen\`eve, 24 quai Ernest Ansermet, 1211 Gen\`eve 4, Switzerland.}


\begin{abstract}
We study braneworlds in a five dimensional bulk, where cosmological
expansion is mimicked by motion through \ads. We show that the five
dimensional graviton reduces to the four dimensional one in the late
time approximation of such braneworlds. Inserting a fixed regulator
brane far from the physical brane, we investigate quantum graviton
production due to the motion of the brane. We show that the massive
Kaluza-Klein modes decouple completely from the massless mode and they
are not generated at all in the limit where the regulator brane
position goes to infinity. In the low energy limit, the massless four
dimensional graviton obeys the usual 4d equation and is therefore also
not generated in a radiation-dominated universe.
\end{abstract}

\pacs{04.50.+h, 11.10.Kk, 98.80.Cq}

\maketitle

\section{Introduction}
In recent years, cosmological models where our Universe is
represented as a hypersurface moving through a higher dimensional
spacetime have received a lot of attention. The reasons for that
are twofold. First, string theory, the presently most promising
candidate of a theory for quantum gravity is consistent only in
more than $3+1$, namely in $9+1$ or (for M-theory) $10+1$
dimensions~\cite{Polchinski:1998rq,Polchinski:1998rr}. Second, the
hierarchy problem {\em i.e.} the unnatural disparity of the
electroweak scale $\sim 1$TeV and the Planck scale $\sim
10^{16}$TeV can be addressed in the context of
extra-dimensions~\cite{Arkani-Hamed:1998rs}. For this, it is
necessary that one or several of the extra-dimensions be much
larger than the Planck scale. The $4$-dimensional effective Planck
mass then becomes $M_4 \simeq \sqrt{M_{4+n}^{2+n} L^n}$, where $n$
denotes the number of extra dimensions and $L$ is their size.
These extra dimensions can be large and nevertheless undetected if
standard model particles are confined to a lower dimensional
hypersurface and cannot probe them. String theory predicts the
existence of such $Dp$--branes onto which standard model particles
are confined~\cite{Polchinski:1995mt}. The bulk spacetime around
the $Dp$-brane can then only be probed by gravity. Gravity has
been tested only down to scales of about $0.1$mm and therefore
allows for $L \lsim 0.1$mm. It has been shown that confinement of
gravity to a region of size $L \lsim 0.1$mm around a brane with
one extra-dimension can also be achieved by a non-compact Anti-de
Sitter spacetime~\cite{Randall:1999vf}. The cosmological situation
of an expanding universe is then obtained by a brane moving
through a $5$-dimensional Anti-de Sitter spacetime (\ads). At low
energy, this setup leads to the usual Friedmann equations for the
expansion of the Universe~\cite{Kraus:1999it,Binetruy:1999hy}. This is the case
which we consider in this paper.

We want to address the following issue: a brane moving through \ads
spacetime leads to time-varying boundary conditions for quantum fields
living in the bulk. In quantum physics it is well known that moving
boundaries yield particle creation from vacuum, so-called motion
induced radiation \cite{Lambrecht:1996un}.  For instance, photons are
produced from vacuum in dynamical cavities (dynamical Casimir effect)
\cite{Dodonov:2001yb} as well as by a single moving mirror
\cite{Maia:1996}.  Here we study the same mechanism in brane world
cosmology, namely graviton generation provoked by the motion of the
brane.  Apart from the massless graviton, braneworlds allow for a
tower of massive Kaluza-Klein gravitons which might also be produced
by the moving brane. Such massive modes have potentially devastating
effects as they would eventually dominate the energy density of the
universe and spoil the phenomenology if their production is
sufficiently copious. Here we shall show explicitly that the dangerous
massive modes are not produced in the single brane case. Also the
production of massless 4-dimensional gravitons is strongly suppressed
at low energy.

In the next section we present our setup and derive the equations of
motion for the graviton modes. We especially discuss the nature of the
coupling matrix which describes the coupling of the different modes
due to the time dependence of the boundary (the brane). The formalism
laid out is very general and we plan to apply it to other setups (high
energy, 2-brane models) in the future. We also show that at low energy
and for a regulator brane placed sufficiently far, the massless
graviton mode obeys precisely the 4d equation of motion for gravitons
in a Friedmann universe and decouples from the massive modes.  In
Section~\ref{s:ana} we brielfly estimate the produced gravitons and
their spectrum analytically, focusing on the dominant 0-mode.  We
summarize our conclusions in Section~\ref{s:con}.

\section{The mode equations for gravitons in moving braneworlds}
\subsection{The background}
We consider the cosmological Randall-Sundrum-II model where we
have \ads geometry in the bulk with one physical brane at a
time-dependent position $\ybr(t)$. In Poincar\'e coordinates, the
bulk metric is given by
\begin{equation}\label{e:bulk-metric}
 \ud s^2
 = g_{\si{AB}} \ud x^{\si{A}} \ud  x^{\si{B}}
 = \frac{L^2}{y^2} \left[-\ud t^2 + \delta_{ij} \ud x^i \ud x^j + \ud
 y^2\right]~.
\end{equation}
Capital Latin indices $A,B$ run from $0$ to $4$ and lower case
Latin indices $i,j$ from $1$ to $3$. Four-dimensional indices
running from $0$ to $3$ will be denoted by lower case Greek
letters. $L$ is the \ads curvature radius and is related to the
negative cosmological constant $\Lambda$ by $\Lambda = -6/L^2$ to
solve the bulk Einstein equations
\begin{equation}
 \label{e:einstein}
 G_{\si{AB}} + \Lambda g_{\si{AB}} = 0~.
\end{equation}
We now introduce a brane at $y=\ybr(t)$ and replace the ``left
hand side'', $0<y< \ybr(t)$, of \ads by a second copy of the
``right hand side''. We use the superscripts ``${\si
>}$'' and ``${\si <}$'' for the bulk sides with $y > \ybr$ and $y
< -\ybr$, respectively. In terms of the coordinate $y$, the value
of $y$ decreases continuously from $\infty$ to $\ybr$ and then
jumps to $-\ybr$ over the brane whereafter it continues to
decrease. At the brane position, $\ybr^{\si >}=\ybr(t), ~
\ybr^{\si <} = - \ybr(t)$, the metric function $(L/y)^2$ has a
kink. The Einstein equations at the brane position are singular,
they contain a Dirac--delta function,
\begin{equation}
 G_{AB} +\Lambda g_{AB}
 = \kappafive T_{AB}^{\textrm{brane}}\delta(y-\ybr)~.
 \label{e:G_AB-singular}
\end{equation}
The delta function confines the energy momentum tensor from
Standard Model fields to the brane. To avoid the delta function,
one can integrate Eq.~\eqref{e:G_AB-singular} over the extra
dimension which leads to the so-called Israel-Darmois junction
conditions~\cite{Lanczos:1924,Sen:1924,Darmois:1927,Israel:1966}
at the brane position. These read~\cite{Misner:1970aa}
\begin{align}
 g_{\mu \nu}^{\si >} - g_{\mu \nu}^{\si <} &=0~,
 \label{e:g-continuous} \\
 K_{\mu \nu}^{\si >} - K_{\mu \nu}^{\si <}
 &= \kappafive\left( S_{\mu \nu} - \frac{1}{3} S q_{\mu \nu} \right)
 \equiv \kappafive \widehat{S}_{\mu \nu}~,
 \label{e:K-jump}
\end{align}
where $S_{\mu \nu}$ is the energy-momentum tensor on the brane with
trace $S$, and
\begin{equation}
 \kappafive \equiv 6\pi^2 G_5=\frac{1}{M_5^{3}}~.
\end{equation}
$M_5$ and $G_5$ are the five-dimensional (fundamental) reduced
Planck mass and Newton constant, respectively. $K_{\mu\nu}$ is the
extrinsic curvature of the brane and $q_{\mu\nu}$ is the induced
metric on the brane. The first junction
condition~\eqref{e:g-continuous} simply states that the induced
metric, the first fundamental form,
\begin{equation}
 q_{\mu\nu} = e^{\si{A}}_\mu e^{\si{B}}_\nu g_{\si{AB}}~,
 \label{e:def-firstfund}
\end{equation}
be continuous across the brane. Here the vectors $e_\mu$ are
tangent to the brane. If we parametrize the brane
by coordinates $\left(z^\mu\right)$ and its position in the bulk
is given by functions $\Xbr^{\si{A}}(z^\mu)$, the vectors $e_\mu$
can be defined by
\begin{equation}
 e_\mu^{\si{A}} = \p_\mu \Xbr^{\si{A}}(z) ~.
\end{equation}
Denoting the brane normal by $n$, we have $ g_{\si{AB}}e^{\si{A}}_\mu
n^{\si{B}}=0$. The extrinsic curvature can be expressed purely in
terms of the internal brane
coordinates~\cite{Deruelle:2000yj,Mukohyama:2001yp},
$K=K_{\mu\nu}dz^\nu dz^\mu$, with
\begin{equation}
 K_{\mu \nu}
 = -\frac{1}{2} \left[g_{\si{AB}}
    \left(e^{\si{A}}_\mu  \p_\nu n^{\si{B}}
          + e^{\si{A}}_\nu \p_\mu n^{\si{B}}\right)
  + e^{\si{A}}_\mu e^{\si{B}}_\nu n^{\si{C}}
 g_{\si{AB},\si{C}} \right]~.
 \label{eq:extusefull}
\end{equation}
The link between the extrinsic curvature of the hypersurface and
the brane energy-momentum tensor is established by the second
junction condition~\eqref{e:K-jump}, which replaces the 4d
Einstein equation.

A homogeneous and isotropic brane moving through \ads with brane
position $\ybr(\eta)$ and bulk time given by $\tbr(\eta)$ has the
metric
\begin{equation}
 \ud s^2
 = \frac{L^2}{\ybr^2(\eta)}
   \left[-\left(1- \left(\frac{\ud\ybr}{\ud t}\right)^2 \right)\ud t^2
   +\delta_{ij}\ud x^i \ud x^j\right]
 = a^2(\eta)\left[ -\ud\eta^2 +\delta_{ij}\ud x^i\ud x^j\right]~,
 \label{eq:branemetric}
\end{equation}
where $a=L/\ybr$ is the scale factor and $\eta$ denotes the
``conformal time'' of an observer on the brane,
\begin{equation}
 \ud\eta = \sqrt{1- \left(\frac{\ud\ybr}{\ud t}\right)^2}dt
         \equiv \ga^{-1}\ud t~.
\end{equation}
The brane motion induces the $\gamma$-factor which relates the
(conformal) eigentime $\eta$ of the brane to the (conformal)
coordinate time $t$ of the bulk. From now on an overdot indicates a
derivative w.r.t. conformal time $\eta$ on the brane. The brane normal
is then given by
\begin{align}
 {n}_0 = -\frac{\dybr}{\dtbr}{n}_4~, \qquad
 {n}_4^2 = \left(\frac{L}{\ybr}\right)^2\dtbr^2
                \left(\dtbr^2-\dybr^2\right)^{-1}~.
\end{align}

We consider a homogeneous and isotropic total energy momentum tensor
on the brane, $ S^\mu_\nu =T^\mu_\nu-\tension \delta^\mu_\nu$. Here
$\tension$ is the brane tension and $T^\mu_\nu$ is the energy momentum
tensor of particles and fields confined on the brane given by $T^0_0 =
-\rho$, $T^i_j = P\delta^i_j$. The second junction conditions now
become
\begin{align}
 \kappafive({\rho}+\tension)
 &= 6\frac{\sqrt{1+L^2H^2}}{L}  ~, \label{e:hub1} \\
 \kappafive({\rho}+ {P})
 &= -\frac{2L\dot{H}}{a\sqrt{1+L^2H^2}} ~, \label{e:ij}\\
 \dot{{\rho}} &=-3Ha({\rho}+P)~,\label{e:T:continuity}\\
 H^2
 &= \frac{\kappafive^2}{18}\tension{\rho}
 \left(1+\frac{{\rho}}{2\tension}\right)
 +\frac{\kappafive^2\tension^2}{36}-\frac{1}{L^2}~,\label{e:T:Hubble}
\end{align}
where $H\equiv \dot a/a^2$. Equations~\eqref{e:hub1} to
\eqref{e:T:Hubble} form the basis of brane cosmology and have been
discussed at length in the literature (for a review,
see~\cite{Maartens:2003tw} or~\cite{Durrer:2005dj}). The last equation
is called the ``modified Friedmann equation'' for brane
cosmology~\cite{Binetruy:1999hy}. For usual matter with $\rho+P>0$,
$\rho$ decreases during expansion and at sufficiently late time
$\rho\ll \tension$. The ordinary $4$-dimensional Friedmann equation is
then recovered if we set
\begin{equation}
 \frac{\kappafive^2 \tension^2}{12}- \frac{3}{L^2} = \Lambda_4
 \quad\text{and}\quad
 \kappareduct =8\pi G_4 = \frac{\kappafive^2 \tension}{6}~.
 \label{e:fine}
\end{equation}
Neglecting the 4-dimensional cosmological constant, $\Lambda_4 \simeq
0$, we obtain in addition
\begin{equation}
 \label{e:fivefour}
 L = \frac{6}{\kappafive \tension}  \quad\text{and}\quad \kappareduct
 =\frac{\kappafive}{L}~. 
\end{equation}
Note that, although for a de Sitter brane the  density is simply
constant, $\rho=-P=\cst$ and there is no late time approximation,
Eq.~\eqref{e:ij} implies that the Hubble rate remains constant and we
reproduce the usual exponential expansion. Only the relation between
the expansion rate $H$ and the brane density $\rho$ is modified.

\subsection{Tensor perturbations}
A quantum field in the bulk is generically expected to be modified
by the moving brane which acts as a moving boundary of the
$5$-dimensional bulk spacetime. We want to study this effect for
bulk gravitons. With this in mind,
we now linearly perturb the bulk metric allowing for tensor
modes\footnote{{}The word ``tensor'' here has to be understood as
spin 2 degree of freedom w.r.t. rotations of the 3-dimensional
homogeneous and isotropic slices normal to the extra-dimension
$y$.},
\begin{align}
 \ud s^2 = \frac{L^2}{y^2}
 \left[-\ud t^2+(\delta_{ij}+2h_{ij})\ud x^i \ud x^j+\ud y^2 \right]~.
\end{align}
Tensor modes satisfy the traceless and transverse conditions, $h_i^i =
\p_ih^i_j = 0$. We then decompose $h_{ij}$ into spatial Fourier modes,
\begin{equation}
 h_{ij}(t,\ve{x},y)
 = \int \frac{d^3k}{(2\pi)^{3/2}} \sum_{\bb=+,\times}
  e^{i\ve{k}\cdot\ve{x}}e_{ij}^{\bb}\Hb(t,y;k)~,
\end{equation}
where $e_{ij}^{\bb}$ are unitary constant transverse-traceless
polarization tensors which form a base of the two polarization
states $\bb = +$ and $\bb = \times$. The perturbed Einstein
equations now yield the equation of motion for the mode function
$\Hb$, which obeys the Klein-Gordon equation for minimally coupled
massless scalar fields in
\ads~\cite{Hawking:2000kj,Hawking:2000bb,Langlois:2000ns}
\begin{equation}
 \left[\p_t^2 +k^2 -\p_y^2 + \frac{3}{y}\p_y \right] \Hb(t,y;k) = 0~.
\label{e:T-bulk-eq}
\end{equation}
In addition to the bulk equation of motion the modes also satisfy
a boundary condition at the brane coming from the second junction
conditions,
\begin{equation}
 \left.\left[LH\p_t \Hb -\sqrt{1+L^2H^2}\p_y\Hb\right]\right|_{\ybr}
 = \left.\ga^{-1}\left(\vv\dd_t -\dd_y\right)\Hb \right|_{\ybr}
 = \frac{\kappafive}{2}aP\PiTb~.
 \label{e:T-JC-general}
\end{equation}
Here $\PiTb$ denotes possible anisotropic stress perturbations in
the brane energy-momentum tensor and we have introduced the brane
velocity $\vv \equiv \frac{LH}{\sqrt{1+L^2H^2}}$. For simplicity
we set $\PiTb =0$ in this work. (Some of the difficulties which appear
when $\PiTb \neq0$ are discussed in~\cite{CR}. The wave
equation~\eqref{e:T-bulk-eq} with boundary
condition~\eqref{e:T-JC-general} cannot be solved analytically
except if the background metric functions are separable, and this
only happens for maximally symmetric branes, i.e., branes with
constant Hubble rate. This includes the Randall-Sundrum case $H =
0$. A cosmologically relevant case is the de-Sitter brane, $0 < H
=\cst$.  The spectrum of gravitational waves generated during de
Sitter brane inflation can be
calculated~\cite{Langlois:2000ns,Gorbunov:2001ge,Frolov:2002qm,
Kobayashi:2003cn}.

\subsection{Late time approximation}
We want to investigate late time and see whether we have graviton
production also at late time, after inflation. We therefore
consider the limit
\begin{equation}
 \vv \equiv \frac{LH}{\sqrt{1+L^2H^2}}\ll 1 ~.
 \label{e:latetime}
\end{equation}
In this limit the boundary condition~\eqref{e:T-JC-general}
reduces to
\begin{equation}
\left. \p_y\Hb\right|_{\ybr} = 0~.
\label{e:T-JC-fixed}
\end{equation}

If the position of the brane is fixed, the solutions of the system
formed by Eq.~\eqref{e:T-bulk-eq} and Eq.~\eqref{e:T-JC-fixed} are
well known. These are the Bessel functions,
\begin{equation}
\Hb = \TNa\exp{\pm i\omega t}
(\TKK y)^2\left[J_2(\TKK y)+\TNb Y_2(\TKK y)\right]  ~,
 \label{e:Hbulk}
\end{equation}
where $\omega =\sqrt{\TKK^2+k^2}$. The junction
condition~\eqref{e:T-JC-fixed} requires
\begin{equation}
\TNb = -\frac{J_1(\TKK \ybr)}{Y_1(\TKK \ybr)}
\simeq \frac{\pi}{4}(\TKK \ybr)^2 ~,
\end{equation}
where the last expression is a good approximation for $\TKK\ybr\ll
1$. This is precisely the result of Randall and
Sundrum~\cite{Randall:1999vf} for a static \ads brane. Since the extra
dimension $y$ is not compact, the mass-spectrum is continuous,
$\TKK^2$ can take any non-negative value. 

Allowing for a single moving brane is not well suited for a numerical
treatment. We therefore introduce a second, so-called ``regulator
brane'' far away from the physical brane, at the position
$y=\yreg$ which we let tend to infinity at the end of our
calculation.  We assume the regular brane to be empty and fixed.
The boundary condition at the regulator brane is thus
\begin{equation}
 \left.\p_y \Hb\right|_{\yreg}= 0~.
\label{e:T-JC-reg}
\end{equation}
For a solution of the form~\eqref{e:Hbulk} this implies the
additional constraint
\begin{equation}
 J_1(\TKK \yreg) +\TNb Y_1(\TKK \yreg) =0 ~.
\label{e:T-JC-regBessel}
\end{equation}
This condition is satisfied only for a discrete series of mass
eigenvalues $\TKK_\alpha$. With $\TNb$ also $\TKK_\alpha$ depends on
the position of the physical brane and therefore on time.

The evolution equation~\eqref{e:T-bulk-eq} together with the boundary
conditions form a Sturm-Liouville problem, with eigenvalue equation
\begin{equation}
 \left[-\p_y^2+\frac{3}{y}\p_y\right]\phi_\alpha
 =-y^3\p_y\left[y^{-3}\p_y \phi_\alpha \right]
 =\TKK_\alpha^2\phi_\alpha~.
\label{e:T-bulk-slow-moving-homogeneous-EV}
\end{equation}
At any given time, Eq.~\eqref{e:T-bulk-slow-moving-homogeneous-EV}
yields an orthonormal system of ``instantaneous solutions''
\begin{align}
 \phi_0 &= \TNa_0 +\TNb_0y^4~, \\
 \phi_i &= \TNa_i(\TKK_i y)^2[J_2(\TKK_i y) + \TNb_i Y_2(\TKK_i y)]
         \equiv \TNa_i(\TKK_i y)^2 C_2(\TKK_iy)~,
\end{align}
where we have introduced $C_\nu(\TKK_iy) \equiv J_\nu(\TKK_iy) +
\TNb_i Y_\nu(\TKK_i y)$. Here and below we use indices $i$ and $j$
to denote all the massive modes and indices $\alpha$ and $\beta$
to denote all the modes, including the massless mode. The boundary
conditions~\eqref{e:T-JC-fixed} and~\eqref{e:T-JC-reg} then
require
\begin{align}
 \TNb_0 = 0 \quad\text{and}\quad
 C_1(\TKK_i\ybr) = C_1(\TKK_i\yreg) = 0~.
 \label{e:T-BC}
\end{align}
Furthermore, the solutions form an orthonormal system of functions
on the Hilbert space $\Hilb =\Leb^2\left([\ybr,\yreg], y^{-3}\ud
y\right)$
\begin{equation}
 \scal{\phi_\alpha}{\phi_\beta}
 \equiv \int_{\ybr}^{\yreg} y^{-3} \phi_\alpha (t,y) \phi_\beta(t,y) \ud y
 =  \delta_{\alpha\beta}~.
 \label{e:T:inner-product}
\end{equation}
or, explicitly,
\begin{align}
 \scal{\phi_i}{\phi_j}
 &= \frac{\TNa_i\TNa_j\TKK_i^2\TKK_j^2}{\TKK_i^2-\TKK_j^2}
    \left.\left[\TKK_j y C_2(\TKK_i y)C_1(\TKK_j y)-\TKK_i y C_1(\TKK_i y)C_2(\TKK_j y)
    \right]\right|_{\ybr}^{\yreg}
  =0~,
  \label{e:T:orthogonality-mn}\\
 \scal{\phi_i}{\phi_0}
 &= -\TNa_i\TKK_i \left.\left[y^{-1}
   \left\{\TNa_0 C_1(\TKK_i y)-\TNb_0y^4C_3(\TKK_i y)\right\}\right]
\right|_{\ybr}^{\yreg}
  =0~,
  \label{e:T:orthogonality-m0}\\
 \scal{\phi_i}{\phi_i}
 &= \frac{1}{2}\left(\TNa_i\TKK_i\right)^2
    \left.\left[(\TKK_i y)^{2} \left\{C_2^2(\TKK_i y)-C_{1}(\TKK_i y)C_{3}(\TKK_i y)
    \right\}\right]\right|_{\ybr}^{\yreg}
  =1~,
  \label{e:T:normalisation-mnApp}\\
 \scal{\phi_0}{\phi_0}
 &= \TNa_0^2 \frac{\yreg^2-\ybr^2}{2\yreg^2\ybr^2}
    +\TNa_0 \TNb_0 (\yreg^2-\ybr^2)+\frac{1}{6}\TNb_0^2 (\yreg^6-\ybr^6)
  =1~.
  \label{e:T:normalisation-m0App}
\end{align}
The orthogonality relations are trivially satisfied with the boundary
conditions~\eqref{e:T-BC}, whereas the normalization conditions fix
the constants $\TNa_\alpha$,
\begin{align}
 \frac{1}{2}\left(\TNa_i\TKK_i\right)^2
  \left. (\TKK_i y)^{2}C_2^2(\TKK_i y) \right|_{\ybr}^{\yreg}   &=1~,
  \label{e:T:normalisation-mn}\\
 \TNa_0^2 \frac{\yreg^2-\ybr^2}{2\yreg^2\ybr^2}  &=1~.
  \label{e:T:normalisation-m0}
\end{align}
All the constants depend over the brane position $\ybr(t)$ on
time. Since the orthonormal set
$\left\{\phi_\alpha(t,y)\right\}_{\alpha=0}^\infty$ is complete in
$\Hilb$, we can expand a generic solution in the form
\begin{equation}
 \Hb(t,y;k) = \sum_{\alpha=0}^\infty \TTF_\alpha(t;k)
 \phi_\alpha(t,y) ~,\qquad
 \TTF_\alpha(t;k) = \scal{\phi_\alpha}{\Hb}~.
 \label{e:T-Fourier-series}
\end{equation}
This enables us to write the second order action in $\Hb$ leading
to the equation of motion for $\Hb$ as an action for the canonically
normalized coefficients $\TF_\alpha$. Denoting the second order
perturbation of the gravitational Lagrangian by $\delta( \sqrt{g}R)$ we
obtain for the action
\begin{align}
 \mathcal{S}(k)
 &=  \frac{1}{2\kappafive}\int \ud t \ud y\ \delta(\sqrt{g}R) \nonumber \\
 &= \frac{L^3}{2\kappafive} \int \ud t \ud y\, y^{-3}
    \left[(\p_t \Hb)^2-(\p_y \Hb)^2-k^2 \Hb^2\right]
    \nonumber \\
 &= \frac{1}{2} \int \ud t \sum_{\alpha,\beta}
    \left[(\p_t \TF_\alpha)(\p_t \TF_\beta)\delta_{\alpha\beta}
          +2M_{\alpha\beta}\TF_\alpha(\p_t \TF_\beta)
          +\left\{N_{\alpha\beta}-\scal{\p_y \phi_\alpha}{\p_y \phi_\beta}
           -k^2 \delta_{\alpha\beta}\right\}\TF_\alpha \TF_\beta \right]
   \nonumber \\
 &= \frac{1}{2} \int \ud t \sum_{\alpha}
    \Big[(\p_t \TF_\alpha)^2-\omega_\alpha^2\TF_\alpha^2
          +\sum_{\beta}\left\{2M_{\alpha\beta}\TF_\alpha \p_t\TF_\beta
          +N_{\alpha\beta} \TF_\alpha \TF_\beta \right\}\Big]~,
   \label{e:T-action-1}
\end{align}
where we have introduced coupling matrices
\begin{equation}
 M_{\alpha\beta}(t)\equiv \scal{\p_t \phi_\alpha}{\phi_\beta}~, \qquad
 N_{\alpha\beta}(t)\equiv \scal{\p_t \phi_\alpha}{\p_t \phi_\beta}~,
 \label{e:T-def-Mmn-Nmn}
\end{equation}
and the canonically normalized coefficients
\begin{equation}
\TF_\alpha = \sqrt{\frac{L^3}{\kappafive}}\,\TTF_\alpha ~. \label{e:canon}
\end{equation}
As before, the frequency is given by
\begin{equation}
\omega_\alpha^2(t;k)\equiv k^2+\TKK_\alpha^2~.
\end{equation}
The equations of motion for the Fourier coefficients $\TF_\alpha(t;k)$
then simply follow from the Euler-Lagrange equations for the
action~\eqref{e:T-action-1},
\begin{equation}
 \p_t^2 \TF_\alpha + \omega_\alpha^2 \TF_\alpha
 + \sum_\beta \left[ M_{\beta\alpha} - M_{\alpha\beta}\right]\p_t\TF_\beta
 + \sum_\beta \left[\p_t M_{\alpha\beta} - N_{\alpha\beta}\right]
 \TF_\beta =0~.
 \label{e:T:Fourier-evol-eq}
\end{equation}

Since $ \scal{\phi_\alpha}{\phi_\beta} =$ constant, one might naively
expect that $ M_{\alpha\beta}$ be antisymmetric. But this is not the
case, since $ M_{\alpha\beta}$ depends on time not only via
$\phi_\alpha$ and $\phi_\beta$, but also via the integration boundary
which enters the definition of the inner
product~\eqref{e:T:inner-product}. On the other hand, the completeness
of the eigenfunctions $\phi_\alpha$ implies
\begin{equation}
 \sum_\gamma \phi_\gamma(y)\phi_\gamma(\tilde{y}) =
 \delta(y-\tilde{y})y^3~,
\end{equation}
so that
\begin{align}
 \sum_\gamma M_{\alpha\gamma}M_{\beta\gamma}
 = \sum_\gamma \left[\scal{\p_t\phi_\alpha}{\phi_\gamma}
                     \scal{\p_t\phi_\beta}{\phi_\gamma}\right]
 = \scal{\p_t\phi_\alpha}{\p_t\phi_\beta} = N_{\alpha\beta}~.
\end{align}

The system of coupled second-order differential
equations~\eqref{e:T:Fourier-evol-eq} thus solely depends on the
time-dependent frequency $\omega_\alpha^2$ and the coupling matrix
$M_{\alpha\beta}$. Numerical solutions of this system in other
situations (oscillating cavities) have been obtained in
Refs.~\cite{Ruser:2004,Ruser:2005xg}. Also the results obtained in
this work are tested with the code described
in~\cite{Ruser:2005xg}. We plan to apply this numerical technique in
future work.

The matrix elements are given in terms of the instantaneous solutions
$\phi_\alpha$. The use of several identities of Bessel functions
finally leads to
\begin{align}
 M_{00} =&~\ybrR \frac{\yreg^2}{\yreg^2-\ybr^2}~,\\
 M_{0j} =&~ 0~, \\
 M_{i0} =&~ 2\sqrt{\TKKR_iM_{00}}~,\\
 M_{ii} =&~ \TKKR_i~,\\
 M_{ij} =&~-\TNa_i\TNa_j\TKKR_i\TKK_i^3\TKK_j^2 \int_{\ybr}^{\yreg} y^2 C_1
            (\TKK_i y)C_0(\TKK_j y)\ud y  \nonumber \\
         &~-\frac{2\TKK_i^2}{\TKK_i^2-\TKK_j^2} \sqrt{\TKKR_i\TKKR_j}
           \left[1+\frac{\TKKR_i}{\ybrR}\right]
 \left[\sqrt{\frac{\TKKR_i(\ybrR+\TKKR_j)}{\TKKR_j(\ybrR+\TKKR_i)}}-1\right]~,
    \label{e:T:Mij-exact}
\end{align}
where we have set
\begin{align}
 \TKKR_i
  & \equiv \p_t \ln{\TKK_i}
  = \ybrR \left[\frac{Y_1^2(\TKK_i\ybr)}{Y_1^2(\TKK_i\yreg)}-1\right]^{-1}
  \qquad\text{and}\\
\ybrR(t)  & \equiv \p_t\ln{\ybr} = -\vv \ybr^{-1} \simeq -Ha =
  -\frac{\dot a}{a} \equiv -\HH ~.
\end{align}
Numerics indicate that both terms appearing in $M_{ij}$ have
similar amplitudes, thus the second term can give us an order of magnitude
estimate of Eq.~\eqref{e:T:Mij-exact}. We are mainly interested in
the limit $\yreg \rightarrow \infty$. Setting $\ep =\ybr/\yreg$ we
find $\TKKR_i \simeq \ybrR\ep^2$ for $\ep\ra 0$. To lowest order in
$\epsilon$ and in the limit $\vv\ll 1$ the matrix elements are given by
\begin{align}
 M_{00} & =  -\HH[1 + \mathcal{O}(\epsilon)]~,\\
 M_{0j} & =  0~, \\
 M_{i0} & =  \HH \mathcal{O}(\epsilon) ~,\\
 M_{ii} & = \HH \mathcal{O}(\epsilon^2)~,\\
 M_{ij} & = \HH\mathcal{O}(\epsilon^2)~.
\end{align}

This leads to the following interesting result. At late times $t\simeq
\eta$ and in the 1-brane limit $\epsilon\rightarrow 0$, the
Kaluza-Klein modes with non-vanishing mass evolve trivially and only
the massless mode is coupled to the brane motion\footnote{{}It is easy
to see that $\ep^{-1}\sum_\alpha M_{\alpha j}$ is bounded for all
values of $\ep$ and therefore we do not have to fear that the infinite
sum might contribute in the limit $\ep\ra 0$.}
\begin{align}  \label{e:ddotq0}
 \ddot\TF_0 + \left[k^2 -\dot\HH -\HH^2\right] \TF_0 &=0~, \\
 \ddot\TF_i + \left[k^2+ \TKK_i^2\right] \TF_i &=0~,
 \quad\text{for}\quad i\neq 0~. \label{e:ddotqi}
\end{align}
Therefore bulk tensor perturbations $\Hb \simeq {\Hb}_0 =
\sqrt{2}(L/a)\TTF_0$ satisfy the equation
\begin{equation}
 \ddHb + 2\HH\dHb + k^2\Hb = 0~.
\end{equation}
This equation is valid everywhere in the bulk, and in particular on
the physical brane where it reproduces exactly the equation for a
$4$-dimensional gravity wave in a Friedmann universe. This explicitly
proofs that at low energy (late times) the homogeneous tensor
perturbation equation in brane cosmology reduces to the
$4$-dimensional tensor perturbation equation. The massive Kaluza-Klein
modes decouple completely from the massless mode in the $1$-brane
limit. The friction term arises from the time-dependent boundary
condition, hence from the motion of the brane, and is not due to the
excitation of the massive modes as claimed in~\cite{Koyama:2004cf}.

This is our first main result. In the cosmological 1-brane model
(RSII with a moving brane), the 4-dimensional graviton obeys the usual
equation of motion and the (canonical) massive modes
behave like   massive particles in Minkowski space. Contrary to the
massless mode, they are not affected by the motion of the brane. 
However note that $k$ and $\TKK$ are the co-moving momentum and mass. 

\section{Quantum graviton production}\label{s:ana}
So far, we have considered the classical problem of solving a partial
differential equation with time-dependent boundary conditions. Now we
go on to consider the quantum production of gravitons. Let us assume
(for the sake of simplicity) that the physical brane $\ybr$ has been
at rest up to some initial time $t_\mr{in}\simeq \eta_\mr{in}\geqslant
L/a_\mr{in}$, and that at this initial time there have been no
gravitons. The graviton field has been in the vacuum state. Now the
brane starts moving and (as seen from an observer on the brane) the
universe expands. Most probably, at some later time $t_\mr{fin}$, the
graviton field will no longer be in the vacuum state, gravitons have
been generated.

Of course since the expansion of a radiation-dominated universe is
decelerating already prior to  $t_\mr{in}$ the universe is more likely to
expand faster than to be at rest, hence our initial conditions are
not realistic. 
To study the quantum generation of gravitons we now consider the
field $\Hb(k)$. In addition to Eq.~\eqref{e:T-Fourier-series}, we
have a similar mode decomposition for its canonical momentum,
\begin{equation}
 \Pib(t,y;k) = \sqrt{\frac{L^3}{\kappafive}}\frac{\p\LL}{\p (\p_t\Hb)}
 =  \sqrt{\frac{L^3}{\kappafive}}\sum_{\alpha=0}^\infty
 \tilde\TM_\alpha(t;k)\phi_\alpha(t,y)= \sum_{\alpha=0}^\infty
 \TM_\alpha(t;k)\phi_\alpha(t,y)~. 
\end{equation}
The last equality defines the expansion coefficients $\TM_\alpha$
of the momentum $\Pib$. In quantizing the field, we promote both
the field $\Hb$ and its conjugate momentum $\Pib$ to operators. We
then impose the equal time canonical commutation relations for
bosonic fields,
\begin{align}
 \com{\QHb(t,y;k)}{\QHb(t,y';k)} &=0~, \\
 \com{\QPib(t,y;k)}{\QPib(t,y';k)} &=0~, \\
 \com{\QHb(t,y;k)}{\QPib(t,y';k)} &=i\delta(y-y')~.
\end{align}
The operator valued coefficients $\TQF_\alpha, ~\TQM_\alpha$ thus
satisfy
\begin{align}
 \com{\TQF_\alpha(t;k)}{\TQF_\beta(t;k)} &=0~, \\
 \com{\TQM_\alpha(t;k)}{\TQM_\beta(t;k)} &=0~, \\
 \com{\TQF_\alpha(t;k)}{\TQM_\beta(t;k)} &= i\delta_{\alpha\beta}~.
\end{align}
As usual in quantum field theory, we work in the Heisenberg
picture. The operator valued coefficients $\TQF_\alpha(t;k)$ still
satisfy the classical equation of
motion~\eqref{e:T:Fourier-evol-eq}. But now we also know the
initial state. Before $t_\mr{in}$ the coefficient operators are
given by
\begin{equation}
\TQF_\alpha(t;k)
 = \frac{1}{\sqrt{2\omega_\alpha}}\left[\hat
    a_\alpha e^{-it\omega_\alpha} +\adj{\hat a}_\alpha
    e^{it\omega_\alpha} \right] \quad\text{for}\quad t<t_\mr{in} ~.
\end{equation}
Here $\TQa_\alpha(k)$ and $\TQc_\alpha(k)$ are the annihilation and
creation operators associated with the initial vacuum state. At time
$t$, we define the vacuum state $\ket{0,t}$ as the state annihilated
by all annihilation operators, $\TQa_\alpha(k,t)\ket{0,t}=0$,
$\forall\,k$ and $\alpha$. 

We now consider a fixed wavenumber $k$, as before we work in the late
time limit, hence $\eta \simeq t$ and we assume a barotropic fluid
$P=w\rho$ with $w=\cst$ on the brane, so that the scale factor is $a
\simeq (\eta/L)^{2/(1+3w)}$. We are mainly interested in a
radiation-dominated universe, $w=1/3$. In the limit where
$\epsilon\rightarrow 0$, we may neglect the intermode coupling.  The
massive Kaluza-Klein modes then obey the massive wave
equation~\eqref{e:ddotqi} with constant mass, hence they do not
experience any particle generation. The massless mode evolves
according to equation
\begin{equation}
 \ddot\TF_0 +
 \left[k^2-\left(\nu^2-\tfrac{1}{4}\right)\eta^{-2}\right]\TF_0= 0~,
 \qquad  \nu = \frac{3(1-w)}{2(1+3w)}~.
 \label{e:T:EOM-pump-field-bis}
\end{equation}
This is a standard harmonic oscillator equation with a time-dependent
mass term $m^2=-\left(\nu^2-\tfrac{1}{4}\right)\eta^{-2} =
\dot\HH +\HH^2 =\ddot a/a$.  In a radiation-dominated universe
$\nu=1/2$, the mass vanishes identically (the expansion velocity $\dot
a$ is constant) and there is no particle creation. During a matter
dominated universe $m^2(\eta)$ is negative and time-dependent, so
there is particle creation, but one easily estimates that the
produced energy density is of the order $\rho_h(\eta)/\rho_c(\eta)
\simeq \ell_4^2/(a\eta_\mr{eq})^2 \simeq (\ell_4/\tau_\mr{eq})^2 \sim
10^{-112}$, which is vanishingly small. Here $\ell_4
=\sqrt{\kappareduct}$ denotes the 4-dimensional Planck scale and
$\tau_{eq}$ is the age of the universe at equality.

\section{Conclusion}\label{s:con}
We have shown that in the single brane cosmological
Randall-Sundrum model at late time, no massive gravitons are
produced due to the motion of the brane through \ads. Even
though we have only shown this for the 4-dimensional spin-2 mode,
the same result is expected for the spin-1 ``gravi-photon'' and
the spin-0 ``gravi-scalar'' modes which obey the same equations (they
are the three additional helicity states of a massive spin-2
particle). Also the massless gravi-photon and gravi-scalar are not
produced in the single brane limit since they become
non-normalizable. The only particle which can be produced in this
model is the massless spin-2 graviton.

At late time, the massless spin-2 graviton obeys exactly the
4-dimensional equation of motion. Therefore, as in 4-dimensional
cosmology, there is no graviton
production during the radiation-dominated phase of the universe. 
This is a very general result in 4-dimensional cosmology:
during the radiation era, minimally coupled
massless particles are not generated, however, if the universe obeys a
different expansion law they are.

Models with two branes which approach each other and later move
apart again (e.g., the ekpyrotic model), might be more interesting for
the study of graviton 
production. Due to mode coupling, we expect particle production also
for a radiation-dominated expansion law in this case. Furthermore,
 in two brane models also massive modes
are produced, which can lead to stringent constraints, as
they soon come to dominate the universe. Unfortunately, the high
energy regime, where most of the particle production takes place,
depends on the details of the model under consideration. We will
discuss the two brane case in a forthcoming paper~\cite{inprep}.

\begin{acknowledgments}
The authors acknowledge support from the Swiss National Science 
Foundation and the Fondation Marc Birkigt.  
\end{acknowledgments}


\end{document}